\documentclass[twocolumn,pre,showpacs,a4paper,aps]{revtex4}
\usepackage{epsfig}
\usepackage{amsmath}
\usepackage{dcolumn}
\usepackage{bm}
\begin{document}

\title{\bf Uncertainty Relations in Statistical Mechanics: 
a numerical study in small lattice systems}
\author{Gen\'is Torrents}
%
%
\affiliation{ Departament d'Estructura i Constituents de la Mat\`eria,
  Universitat de Barcelona. Diagonal 647, Facultat de F\'{\i}sica,
  08028 Barcelona, Catalonia}
\author{Eduard Vives}
%
%
\affiliation{ Departament d'Estructura i Constituents de la Mat\`eria,
 Universitat de Barcelona. Diagonal 647, Facultat de F\'{\i}sica,
  08028 Barcelona, Catalonia}
\author{Antoni Planes}
%
%
%
\affiliation{ Departament d'Estructura i Constituents de la Mat\`eria,
  Universitat de Barcelona. Diagonal 647, Facultat de F\'{\i}sica,
  08028 Barcelona, Catalonia}
\date{\today}

\begin{abstract}
We  have analyzed the  validity of  uncertainty relations  between the
fluctuations of  thermodynamically conjugated extensive  and intensive
variables  within  the field  of  statistical  mechanics. Analysis  is
presented for  two particular examples  of small lattice  systems that
are  in  contact  with   reservoirs  of  comparable  sizes:  an  Ising
paramagnet  and an  Ising ferromagnet.   The numerical  results enable
determination of the range  of applicablity of the proposed relations.
Due to the  fact that the examples correspond  to systems described by
discrete  variables, the uncertainty  relations are  not valid  if the
probability of boundary states is too high.
\end{abstract}


\maketitle

\section{\label{sec:Intro} Introduction}

Uncertainty  relations  can  be  established in  any  physical  theory
involving  random conjugated  variables  characterized by  probability
distribution  functions.  The  present paper  deals  with uncertainty
relations in statistical mechanics.

Statistical  mechanics  combines  equilibrium  thermodynamics  with  a
probability  postulate  \cite{TerHaar1955,Huang1987}.   Following  the
standard Gibbs  ensemble theory, in the microcanonical  ensemble it is
assumed   that  all   mechanical  variables   \footnote{The  so-called
"mechanical"   variables  such   as  energy,   volume,  magnetization,
polarization, etc..., which  are thermodynamic extensive variables for
macroscopic bodies.}  are constrained to a fixed value (or are allowed
to vary  within a fixed  small range), while  their entropy-conjugated
variables  are   free  to  fluctuate.    Other  ensembles  (Canonical,
Grand-Canonical, ...)   are introduced  by relaxing the  constraint on
one or more mechanical variables.  This is done by assuming that these
mechanical  variables can  be  exchanged with  a  reservoir, which  is
defined as a macroscopic body  of dimensions much larger than those of
the system.  The  exchange is supposed to be  controlled by parameters
(which  are characteristic of  the reservoir)  that are  identified as
intensive variables.   Formally, the change of  variables when passing
from the microcanonical to  other ensembles is undertaken via Legendre
transforms. In the spirit of equilibrium thermodynamics, the intensive
variables are assumed to have the same value both in the reservoir and
in the  system and are  usually treated as  non-fluctuating variables.
Therefore, within this basic  framework, uncertainty relations are not
expected to apply \cite{Kittel1988}.

While it  seems reasonable  that intensive variables  of a  very large
reservoir  (a  system characterized  by  large  amounts of  mechanical
variables   \footnote{A  very   large  thermal   reservoir   would  be
characterized by an  infinite amount of energy such  that its state is
not affected  by exchanging any finite  amount of energy  with a small
body}) are  treated as non-fluctuating  variables, there is  no reason
for  such   an  assumption  when  dealing   with  intensive  variables
corresponding  to a  small reservoir  or system.   Therefore,  we will
adopt the point of view that, in general, for a system in contact with
a reservoir  (regardless of size) fluctuations  of intensive variables
should  occur  in  the  system  induced by  the  fluctuations  of  the
corresponding  conjugated mechanical  variables.  From  this  point of
view, uncertainty relations are expected to play an important role for
a correct  understanding of the behaviour  of equilibrium fluctuations
of  thermodynamic  variables, with  consequences  in the  experimental
measurements  of these  variables.  Actually  this simple  picture has
received  some experimental  support  \cite{Chui1992} and  thus it  is
reasonable  to  ask  whether  or  not  uncertainty  relations  can  be
formulated and what their applicability limits are.

The problem  was already discussed by  Bohr and Heisenberg  as soon as
Heisenberg  proposed  the   {\it  uncertainty  principle}  in  quantum
mechanics \cite{Bohr}. Therefore, during  the last century a number of
papers  were  published dealing  with  this  problem.  An  interesting
article where these  issues have been reviewed is  the paper by Uffink
and  van  Lith  \cite{Uffink1999}.   More  recently,  Vel\'azquez  and
Curilef   \cite{Velasquez2009}  revisited   the  problem   within  the
framework    discussed    some     years    before    by    Mandelbrot
\cite{Mandelbrot1962}, Gilmore \cite{Gilmore1985} and others. The vast
majority of these works have been formulated from a purely theoretical
point of view, and  applications to particular examples (analytical or
numerical) are scarce. The aim  of the present paper is to numerically
analyze the  validity of uncertainty  relations in two  simple lattice
models  and   shed  some  light   on controversial  interpretations
\cite{Mandelbrot1962,Kittel1988,Kittel1989}

The paper  is organized as follows. In  section \ref{sec:Relations} we
summarize  a derivation  of the  uncertainty relations  in statistical
mechanics.  In  section \ref{sec:Examples} we  numerically analyze two
examples   corresponding  to   an  Ising   paramagnet  and   an  Ising
ferromagnet.  Finally, in Section \ref{sec:Conclusions} we discuss the
relevance of  the results in  relation to temperature  measurements in
small thermodynamic systems and draw some conclusions.

\section{\label{sec:Relations} Uncertainty relations in
statistical mechanics}

\subsection{\label{subsec:Math}Fluctuations in continuous probability
distributions}

Following    the    mathematical    treatment   proposed    in    Ref.
\onlinecite{Velasquez2009}, let  us consider a  continuous probability
distribution,  $p\left   (\bf{x}\right)d\bf{x}$,  defined  within  the
domain $\Omega \in \Re^n$, which  is assumed to vanish at the boundary
$\partial  \Omega$.   The  average  value  of the  derivative  of  any
analytical function $f\left(\bf{x}\right)$ with  respect to one of the
components of $\bf{x}$, can be expressed as:
\begin{widetext}
\begin{equation}
\begin{split}
\left\langle     \frac{\partial     f}{\partial    x_{i}}\right\rangle
=\underset{\Omega}{\int}d^{n}xp\left(\bf{x}\right)\frac{\partial
f\left(\bf{x}\right)}{\partial                 x_{i}}=\underset{\Omega}
{\int}d^{n}x\left(\frac{\partial}{\partial
x_{i}}\left(p\left(\bf{x}\right)f\left(\bf{x}\right)\right)-
f\left(\bf{x}   \right)\frac{\partial   p\left(\bf{x}\right)}{\partial
x_{i}}\right)=\\=-\underset{\Omega}{\int}d^{n}xp\left(\bf{x}\right)f
\left(\bf{x}\right)\frac{1}{p\left(\bf{x}\right)}\frac{\partial
p\left(\bf{x}\right)}{\partial                 x_{i}}\equiv\left\langle
f\left(\bf{x}\right)\Gamma_{i}\left(\bf{x}\right)\right\rangle\
\label{eq:Result1}
\end{split}
\end{equation}
\end{widetext}
where we have defined the affinities:
\begin{equation}
\Gamma_{i}\left(\bf{x}\right)\equiv-\frac{1}{p(\bf{x})}\frac {\partial
p\left(\bf{x}\right)}{\partial        x_{i}}=-\frac{\partial       \ln
p\left(\bf{x}\right)}{\partial x_{i}}\; .
\label{eq:TheAffinity}
\end{equation}
Two interesting  particular cases of the  result in (\ref{eq:Result1})
can   be    obtained   by   choosing    $f\left(\bf{x}\right)=1$   and
$f\left(\bf{x}\right)=x_{j}$. In the first case one gets
\begin{equation}
\left\langle \Gamma_{i}\right\rangle =0 \; \; \forall i ,
\label{eq:term1}
\end{equation}
and in the second case
\begin{equation}
\left\langle x_{j}\Gamma_{i}\right\rangle =\delta_{ij}.
\label{eq:term2}
\end{equation}
Let us now define the differences:
\begin{equation}
\Delta x_{j}\equiv x_{j}-\left\langle x_{j}\right\rangle\; ,
\end{equation}
\begin{equation}
\Delta\Gamma_{i}\equiv\Gamma_{i}-\left\langle\Gamma_{i}\right\rangle
\; .
\end{equation}
The  variances  $\left\langle (\Delta  A)^2  \right  \rangle$ and  the
covariances  $\left\langle\Delta A\Delta B\right\rangle$  of arbitrary
differences $\Delta A$  and $\Delta B$ can be  interpreted as a scalar
products  in  a vectorial  space  where the  metric  is  given by  the
probability  distribution.   In particular,  the  square  root of  the
variance $\sqrt{  \langle \left ( \Delta  A \right )^2  \rangle}$ of a
random variable $A$  (which we will denote as  the fluctuation of $A$)
is the  norm of the vector  $\Delta A$ in this  metric.  Therefore, in
our case, taking into account the Schwartz inequality, we can write:
\begin{equation}
\sqrt{\left\langle \left(\Delta x_{j} \right)^{2} \right \rangle \left
\langle\left(\Delta\Gamma_{i}\right)^{2}\right\rangle}\ge\left \langle
\Delta x_{j}\Delta\Gamma_{i}\right\rangle \; .
\end{equation}
From (\ref{eq:term1})  and (\ref{eq:term2}), the  right-hand-side term
can be expressed as:
\begin{equation}
\left\langle   \Delta  x_{j}\Delta\Gamma_{i}\right\rangle=\left\langle
x_{j}\Gamma_{i}\right\rangle                              -\left\langle
x_{j}\right\rangle\left\langle \Gamma_{i}\right\rangle=\delta_{ij}\; .
\end{equation}
Thus, we obtain  an inequality relating the variances  of the variable
$x_i$ and the affinity $\Gamma_j$:
\begin{equation}
\sqrt{\left\langle      \left(\Delta     x_{j}\right)^{2}\right\rangle
\left\langle   \left(\Delta\Gamma_{i}\right)^{2}\right\rangle   }  \ge
\delta_{ij}\;.
\label{eq:Resultatbo}
\end{equation}
 It  is   worth  noting  that   other  authors  have   used  different
mathematical  approaches  \cite{Gilmore1985,Uffink1999,Falcini2011} to
treat this problem.

\subsection{\label{subsec:StatMech} Application to Statistical
Mechanics:}

In this subsection  we will discuss how to  apply the results obtained
above  in  order to  establish  uncertainty  relations in  statistical
mechanics.   Actually, the  key point  will  be to  find a  convenient
definition of the set $\Omega$, the probabilities $p({\bf x})$ and the
affinities $\Gamma_i$.

We proceed by dividing an isolated universe into two interacting parts
that can exchange one or more mechanical variables.  One part (usually
the smallest one, which we wish to study) will be called the "system",
while the other part will be the "reservoir".  For the closed universe
we assume Boltzmann equiprobability  and therefore its entropy will be
given by:
\begin{equation}
{\cal S}/k = \ln{W} \; ,
\label{eq:Boltzmann}
\end{equation}
where $W$ is  the total number of microstates of  the universe and $k$
is the Boltzmann constant.

Let us  consider the set  $\Omega$ of mechanical  variables describing
the  system  ${\bf  X}  = (X_1,X_2,\cdots  X_n)$.   The  corresponding
variables  for  the  universe   ${\bf  X}^U$  are  constant  (closed).
Therefore, for the reservoir we  have ${\bf X}^R = {\bf X}^U-{\bf X}$.
From knowledge  of  the  number of  microstates  of the  universe
$W({\bf X})$ for which the  system variables take values within $({\bf
X}, {\bf X}+d{\bf X})$ we can determine the probability density:
\begin{equation}
p\left({\bf   X}\right)d{\bf   X}=\frac{W\left(\bf{X}\right)d\bf{X}}{W
\Omega_0} \; ,
\label{eq:distribution}
\end{equation}
where  $\Omega_0$ is  an elementary  volume in  $\Omega$ that  will be
irrelevant for further developments. It is introduced here in order to
have consistent units.  Assuming that this probability density $p({\bf
X})$ is a finite  function, Eq.~(\ref{eq:distribution}) can be written
in terms of a  {\it universe dimensionless entropy} which depends
on the system variables:
\begin{equation}
S^U\left(\bf{X}\right)\equiv \ln \left [ W\left(\bf{X}\right)\right ]
\end{equation}
Therefore we have
\begin{equation}
p\left({\bf X}\right)d{\bf  X}=\frac{e^{S^U\left({\bf X}\right)} d{\bf
X}}{W \Omega_0}\; ,
\label{eq:Theprob}
\end{equation}
which  is none other  than Einstein's  formula for  fluctuations
\cite{Einstein1,  Einstein2}.   At  this  point,  without  making  any
assumptions  about  the  separability  of the  function  $S^U\left({\bf
X}\right)$,  we  can define  the  affinities  between  the system  and
reservoir as the partial derivative:
\begin{equation}
\Gamma_{i}=-\frac{\partial S^U\left({\bf X}\right)}{\partial X_{i}}\;.
\label{eq:Afinitat}
\end{equation}
Note that $\Gamma_i$  has units inverse of those  of $X_i$ since $S^U$
is dimensionless. In some cases, it can be assumed that $S^U\left({\bf
X}\right)$  can be  split into  a  part related  to the  entropy of  the
studied system, $S$, and a  part related to the reservoir, $S^{R}$, so
that
\begin{equation}
S^U\left({\bf    X}\right)=S\left({\bf    X}\right)+   S^{R}\left({\bf
X}^{U}-{\bf X}\right)\,.
\end{equation} 
Since functions $S$ and $S^{R}$  only depend on the internal structure
of the corresponding parts, it  then seems natural to define intensive
variables for the system and reservoir in analogy with thermodynamics:
\begin{eqnarray}
\gamma_{i} & = & \frac{\partial     S     \left({\bf     X}\right)}{\partial
X_{i}} \nonumber \\
\gamma^{R}_{i} & = & \frac{\partial   S^{R}\left({\bf   X}^{U}-{\bf
X}\right)}{\partial \left(X^{U}_{i}-X_{i}\right)}\;.
\end{eqnarray}
Therefore, assuming separability, the affinity can be expressed as:
\begin{equation}
\Gamma_{i}=\gamma_{i}^{R}-\gamma_{i}\;.
\label{eq:Afinitat2}
\end{equation}

The results derived in subsection \ref{subsec:Math} can now be written
in terms of the preceding definitions.  Without assuming separability,
from Eq.~(\ref{eq:term1}) we get:
\begin{equation}
\left \langle \Gamma_i \right \rangle=0\;,
\label{eq:Equilibri}
\end{equation}
and  from the  inequality  relation in  (\ref{eq:Resultatbo}), we  can
write:
\begin{equation}
\sqrt{\left\langle      \left(\Delta     X_{j}\right)^{2}\right\rangle
\left\langle \Gamma_i^{2} \right\rangle}\ge\delta_{ij}\;.
\label{eq:Resultatest}
\end{equation}
When   separability    holds,   equations   (\ref{eq:Equilibri})   and
(\ref{eq:Resultatest})become
\begin{equation}
\left\langle \gamma_{i}^{R}-\gamma_{i} \right\rangle =0\
\label{eq:Equilibri2}
\end{equation}
and
\begin{equation}
\sqrt{\left\langle      \left(\Delta     X_{j}\right)^{2}\right\rangle
\left\langle  \left( \gamma_i^R-  \gamma_i \right  )^{2} \right\rangle
}\ge\delta_{ij}\;,
\label{eq:Resultatest2}
\end{equation}
respectively. These are the main results that we will numerically test
in  the examples  in section  \ref{sec:Examples}.  Note  that  for the
particular case  of choosing energy as a mechanical  variable, in
the separable case, the above two equations read:
\begin{equation}
\left\langle \beta^{R}-\beta \right\rangle =0
\label{eq:betes}
\end{equation}
and
\begin{equation}
\sqrt{\left\langle  \Delta E^{2}\right  \rangle  \left\langle \left  (
\beta^{R}-\beta \right)^{2}\right\rangle }\ge 1\;,
\label{eq:energ}
\end{equation}
where we  have used the more  common symbols $\beta$  and $\beta^R$ to
represent  the intensive variables  conjugated for the energy  of the
system and reservoir.

Equations      (\ref{eq:Equilibri}),      (\ref{eq:Equilibri2})     or
(\ref{eq:betes}) are essentially probabilistic versions of the Zero-th
Principle  of  Thermodynamics   \cite{Redlich1970}.   Because  of  the
fluctuations of  the system and reservoir,  the equilibrium conditions
are not an  absolute equality, but only apply  to statistical averages
of the  intensive variables.   It is important  to note that  with the
separability assumption the conjugated variables of a system appear as
functions  of   extensive  variables  of   the  same  system   (as  in
thermodynamics) and that the same happens when the roles of the system
and reservoir are interchanged.

Equations     (\ref{eq:Resultatest}),    (\ref{eq:Resultatest2})    or
(\ref{eq:energ}) state  that fluctuations in  conjugated thermodynamic
variables are  related in a form  which is similar  to the uncertainty
relations in quantum mechanics.  These results are usually referred to
as uncertainty relations in statistical physics and, even though their
interpretation in quantum mechanics  is different, they both come from
the     same    mathematical     base,    as     noted     in    Ref.~
\onlinecite{Gilmore1985}.

At this point it is convenient  to point out that the previous results
were  obtained after  neglecting  the contribution  from the  boundary
$\partial  \Omega$ in Eq.~(\ref{eq:Result1})  and after  considering a
continuous character  for the  variables $X_i$ describing  the system.
Therefore,  the uncertainty  relation  (\ref{eq:Resultatest}) and  the
equilibrium condition (\ref{eq:Equilibri}) may not hold if states with
extreme  values of  the  variables $X_i$  (at  the boundary  $\partial
\Omega$)  are likely  to occur.   Such states  will be  called boundary
states and in the following  examples, that have a discrete character,
it will  be important  to evaluate their  probability and  check under
which circumstances they become too high.
 
\subsection{\label{subsec:FirstOrder} First-order approximation}

Let us  consider a universe  in which the dimensionless  entropy $S^U$
can be separated between the system and reservoir and we focus only on the
energy of the  system $E$ as a mechanical variable.   Let us assume that
$S^U$   or   equivalently,    the   continuous   probability   density
$p\left(E\right)dE$, is  centred about a maximum at  $E^{eq}$. We also
require  that the  function  $S^U$ has  no  other maxima,  which is  a
typical situation in statistical  mechanics, far from phase transition
points.   Exactly  at  the  maximum we  have  $\Gamma=\beta^R-\beta=0$
(otherwise it would not  be extremal).  Fluctuations of $\Gamma$ about
zero  are related to  how much  the system  fluctuates near  the point
$E^{eq}$.

\begin{figure}[htb]
\begin{center}
\epsfig{file=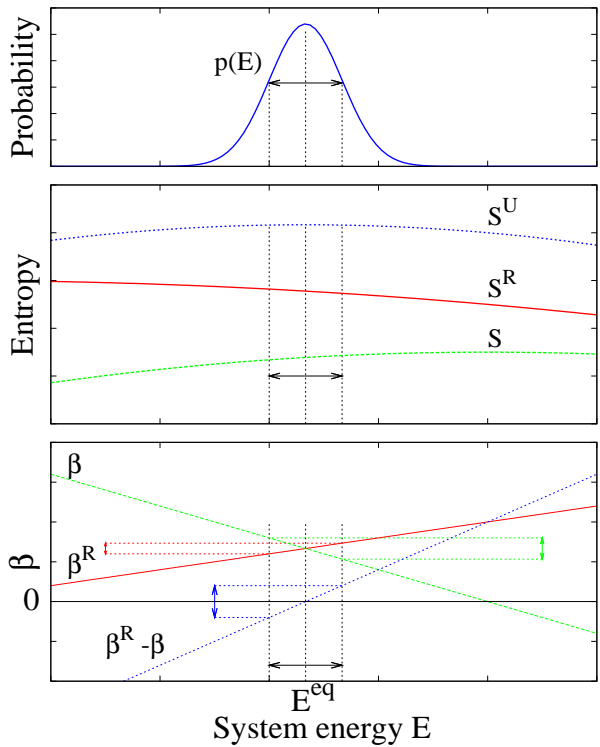,width=8.0cm,clip=}
\end{center}
\caption{\label{FIG1} Schematic representation  of the dependence on
the system energy $E$ of the probability density (above), the entropies
$S^U$, $S^R$ and $S$ (middle), and the inverse temperatures $\beta^R$,
$\beta$,  and the  affinity $\Gamma  = \beta^R  - \beta$  (below). The
arrows indicate the width of the typical fluctuations of the different
variables.}
\end{figure}

Fig.~\ref{FIG1} shows  a schematic representation  of this situation.
The plot above shows the  probability of the energy fluctuations.  The
entropies of the  system, the reservoir and the  universe are shown in
the  middle diagram.  At  the bottom  we show  the derivatives  of the
entropies     corresponding      to     $\beta$,     $\beta^R$     and
$\Gamma=\beta^R-\beta$.   The   arrows  indicate  the   width  of  the
fluctuations  of  the energy  and  the  inverse  temperatures. If  the
probability  density   becomes  narrower,  energy   fluctuations  will
decrease, but $S^U$  will have a more pronounced  maximum.  This means
that the  slope of $\beta^R-\beta$  (blue line in the  bottom diagram)
increases and consequently the affinity will have larger fluctuations.
If energy  fluctuations are  small enough we  can expand  the function
$\Gamma(E)$  about  the point  $E^{eq}$.  A first-order  approximation
gives:
\begin{equation}
\begin{split}
\Gamma   =   \beta^{R}-\beta  \approx\frac{\partial\beta^{R}\left(E^U-
E\right)}{\partial       E}\left(E^{eq}-E\right)-\\-\frac{\partial\beta
\left(E\right)}{\partial E}\left(E-E^{eq}\right) = \phi\Delta E \;,
\end{split}
\label{eq:Linealitz}
\end{equation}
where  we have  taken into  account the fact that  $E^U$ is  fixed and  we have
defined:
\begin{equation}
\phi\equiv   \left  .   \frac{\partial^2  S^U}{\partial   E^2}  \right
|_{E^{eq}}                         =                         -\left.
\frac{\partial\beta^{R}\left(E\right)}{\partial    E}\right|_{E^{eq}}
-\left.\frac{\partial\beta      \left(E\right)}{\partial     E}\right|
_{E^{eq}} \; ,
\end{equation}
which plays  the role of an  inverse joint heat  capacity.  Indeed the
plots  shown   in  Fig.~\ref{FIG1}  correspond   to  such  a first-order
approximation,  where   the  probability  is   Gaussian,  the  entropy
densities  are parabolas  and  the inverse  temperatures and  affinity
behave linearly.

Note   that   within   this   first-order   approximation,   condition
(\ref{eq:betes}) yields  $\langle E  \rangle = E^{eq}$.   Besides, the
fact that $\beta^{R}-\beta  \propto\Delta E$ makes (\ref{eq:energ}) an
equalty (it  is indeed  a property of  the Schwartz  inequality). Thus,
replacing (\ref{eq:Linealitz}) into (\ref{eq:energ}), we find
\begin{equation}
\phi\left\langle \left(\Delta E\right)^{2}\right\rangle =1 \, .
\end{equation}
Hence
\begin{equation}
\left\langle  \left(\Delta E\right)^{2}\right\rangle =\left  [ -\left.
\frac{\partial\beta^{R}\left(E       \right)}{\partial       E}\right|
_{E^{eq}}-\left.\frac{\partial\beta          \left(E\right)}{\partial
E}\right|_{E^{eq}}\right ]^{-1}
\end{equation}

If  we define  the system  heat capacity  $C$ and  the  reservoir heat
capacity $C^R$ as
\begin{equation}
-\left   .   \frac{\partial\beta}{\partial   E}   \right  |_{E^{eq}}
=\frac{1}{kT^{2}}\frac{\partial T}{\partial E}=\frac{1}{kT^{2}C}\;,
\end{equation}
\begin{equation}
-\left  .    \frac{\partial\beta^R}{\partial  E^R}  \right  |_{E^{eq}}=
\left  .    \frac{\partial\beta^R}{\partial  E}  \right  |_{E^{eq}}
=\frac{1}{kT^{2}}\frac{\partial T}{\partial E}=\frac{1}{kT^{2}C^R}\;,
\end{equation}
where $T=\left(k\left\langle\beta\right\rangle\right)^{-1}$ stands for
the equilibrium temperature, we get:
\begin{equation}
\left\langle \left(\Delta  E\right)^{2}\right\rangle = \frac{kT^{2}C^R
C}{C+C^R} \; ,
\label{eq:Lindhart}
\end{equation}
which  is  essentially  the  same  result  obtained  by  Lindhart,  as
discussed in  Ref.~\onlinecite{Uffink1999}, but seen  from within a different
framework.  At the same order of approximation we can compute:
\begin{equation}
\langle   \Gamma^2   \rangle   =  \frac{1}{\left\langle   \left(\Delta
E\right)^{2}\right\rangle}=\frac{C^R+C}{k T^2 C C^R}\;,
\label{eq:Lindhart2}
\end{equation}
\begin{equation}
\langle  \left   (  \Delta\beta^R   \right  )^2  \rangle   =  \frac{C}
{kT^2C^R(C^R+C)}\;,
\label{eq:Lindhart3}
\end{equation}
\begin{equation}
\langle  \left (  \Delta\beta  \right )^2  \rangle =  \frac{C^R}{kT^2C
(C^R+C)}\;,
\label{eq:Lindhart4}
\end{equation}
\begin{equation}
\langle  \left   (  \Delta\beta  \Delta\beta^R  \right   )  \rangle  =
\sqrt{\left(\Delta\beta^{R}\right)^{2}\left(\Delta\beta\right)^{2}}=
\frac{1}{kT^2\left(C^{R}+C\right)}\;.
\label{eq:Lindhart5}
\end{equation}

From the previous equations, for  an infinite reservoir ($C^R \gg C $)
we obtain the expected results
\begin{equation}
\left\langle \left(\Delta E\right)^{2}\right\rangle =kT^{2}C\;,
\end{equation}
\begin{equation}
\langle \Gamma^2 \rangle =\frac{1}{k T^2 C}\;,
\end{equation}
\begin{equation}
\langle \left ( \Delta\beta^R \right )^2 \rangle = 0\;,
\end{equation}
\begin{equation}
\langle \left ( \Delta\beta \right )^2 \rangle = \frac{1}{kT^2C}\;,
\end{equation}
\begin{equation}
\langle\left(\Delta\beta\Delta\beta^R\right)\rangle=0\;.
\end{equation}

\subsection{\label{subsec:Discrete}Applicability to discrete systems}

Even  though many macroscopic  systems can  be studied  as continuous,
their   nature  is   usually  discrete,   and,   therefore,  equations
(\ref{eq:Equilibri})  and  (\ref{eq:Resultatest})  should be  seen  as
approximations. We will now discuss how to deal with discrete systems.

The    main    difficulty    in    the   application    of    equation
(\ref{eq:Resultatest}) to discrete  systems, is how derivatives should
be taken in the definition  of intensive variables.  We must consider,
not only the problem of  finding whether defining forward, backward or
central derivatives can affect the  result, but we must also take into
account  the fact that  $\frac{d}{dx}\ln  p  $  and  $p^{-1}\frac{d}{dx}p$  have
different  corresponding  discrete expressions.   Let  us discuss  two
possible definitions.

If  we  consider  the  second  part  of  Eq.~(\ref{eq:TheAffinity})  a
possible (centred) definition of $\Gamma$ is
\begin{equation}
\Gamma_{i}\equiv-\frac{\ln p_{i+1}-\ln p_{i-1}}{2} \; ,
\label{eq:Afinitatlog}
\end{equation}
which is  separable if  the probability $p$  factorizes.  Nevertheless
this prescription also raises some problems: since the logarithm goes
to  minus infinity as  probability goes  to zero,  one must  be really
careful  at  the   edges  of  the  domain  and   change  the  centred
prescription to backwards and forward prescriptions.

An alternative  definition of $\Gamma_{i}$  is the one  suggested from
the first part of Eq.~(\ref{eq:TheAffinity}):
\begin{equation}
\Gamma_{i}\equiv-\frac{p_{i+1}-p_{i-1}}{2p_{i}} \; .
\label{eq:Afinitatdisc}
\end{equation}
This definition is not separable as  a product of terms for the system
and the  reservoir, not  even when the  probability factorizes.   As a
consequence, the concept of intensive variables cannot be thought of as a
property of a single system. Instead, in this case, we shall deal with
affinities that are a property of each pair of systems in equilibrium.
In the  first example in  the next Section \ref{sec:Examples}  we will
compare the  results obtained with  the two prescriptions in  order to
evaluate how much the choice can affect the results for small systems.

\section{\label{sec:Examples} Numerical simulations}

In order  to discuss  the applicability of  the above  expressions, we
will   test   them   in  two  numerically treatable examples:   a
non-interacting  lattice  system (Paramagnetic  Ising  system) and  an
interacting  lattice  system (Ferromagnetic  Ising  system).  We  will
focus  on   the  validity  of   expressions  (\ref{eq:Equilibri})  and
(\ref{eq:Resultatest})  for  various  universe sizes,  various  system
sizes  within each  universe and  different values  for  the extensive
fixed  variables  of the  universe.   Our  aim is to  deepen  our
understanding of the validity of these equations.

\subsection{\label{subsec:Example1} Paramagnetic Ising system}

Let us  firstly analyze the  validity of the formalism  developed above
for  a  model  without  interaction.   We shall  consider  a  universe
consisting  of $N^U$  non-interacting spins  $s_i$ that  can  take two
values $\pm 1$, depending on whether they are parallel or antiparallel
to an  external field $H$ (which  is a fixed parameter  of the model).
The Hamiltonian is  given by ${\cal H} =  -\sum_{i=1}^N s_i H$.  Since
the universe is isolated (constant energy) the number of spins $N^U_+$
in the  $+1$ state and the number  of spins $N^U_-$ in  the $-1$ state
(fulfilling $N^U_+  + N^U_-=N^U$) are fixed. Therefore,  the number of
possible configurations of the universe is given by:
\begin{equation}
W = \frac{N^U!}{N^U_+! \left ( N^U-N^U_+ \right )!}
\end{equation}
We now consider  a partition of the whole universe  into the two sets:
the system and the reservoir  so that $N^U=N^R+N$.  We will choose the
number of spins $N_+$ in the  $+1$ state in the system as the internal
variable  of  interest in  this  example  (this  choice gives  simpler
expressions than choosing the  energy).  Its probability function will
be given by:
\begin{equation}
\begin{split}
p(N_+)      =      \frac{N!}{N_+!       (N-N_+)!}       \\      \times
 \frac{(N^U-N)!}{(N^U_+-N_+)!    (N^U-N-N^U_+   +N_+)!}    \\   \times
 \frac{N^U_+! \left ( N^U-N^U_+ \right )!}{N^U!} \; ,
\label{probparamagnet}
\end{split}
\end{equation}
where $N^U$,  $N^U_+$ and $N$  are parameters. Note that  the variable
$N_+$ ranges  between the extremes corresponding to  its minimum value
given by  $N_+^{min}=\max(0, N^U_+-N^U+N)$ and its  maximum value given
by   $N_+^{max}=  \min(N^U_+,N)$.    Using  the   probability  function
(\ref{probparamagnet})  we can  explicitly compute  the  average value
$\langle N_+\rangle$ and its variance as
\begin{equation}
\langle \left  ( \Delta  N_+ \right )^2  \rangle = \langle  \left (N_+
\right )^2 \rangle - \langle N_+ \rangle^2
\end{equation}
\begin{figure}[htb]
\begin{center}
\epsfig{file=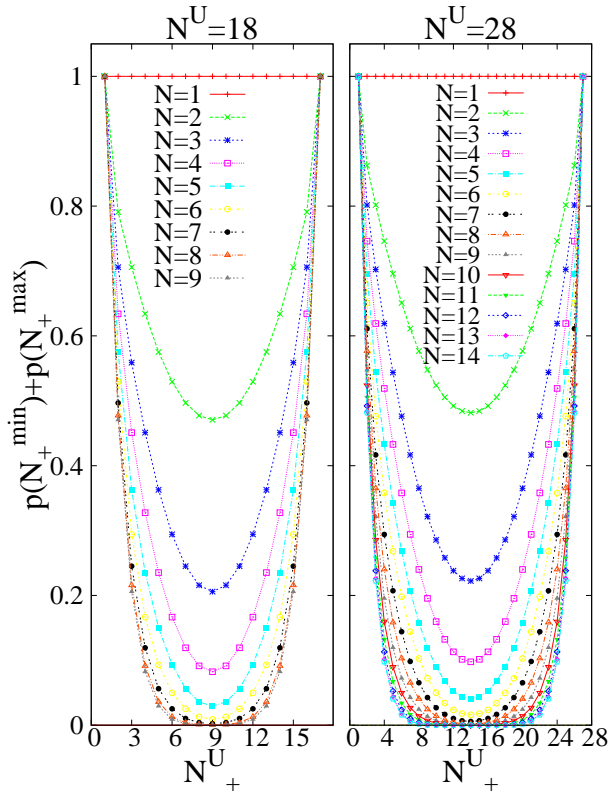,width=8.0cm,clip=}
\end{center}
\caption{\label{FIG2}    Probability    of    the    boundary    states
$p(N_+^{min})+p(N_+^{max})$   as   a  function   of   $N^U_+$  for   a
paramagnetic system  for different universe sizes and  system sizes as
indicated by the title and legend respectively.}
\end{figure}
\begin{figure}[htb]
\begin{center}
\epsfig{file=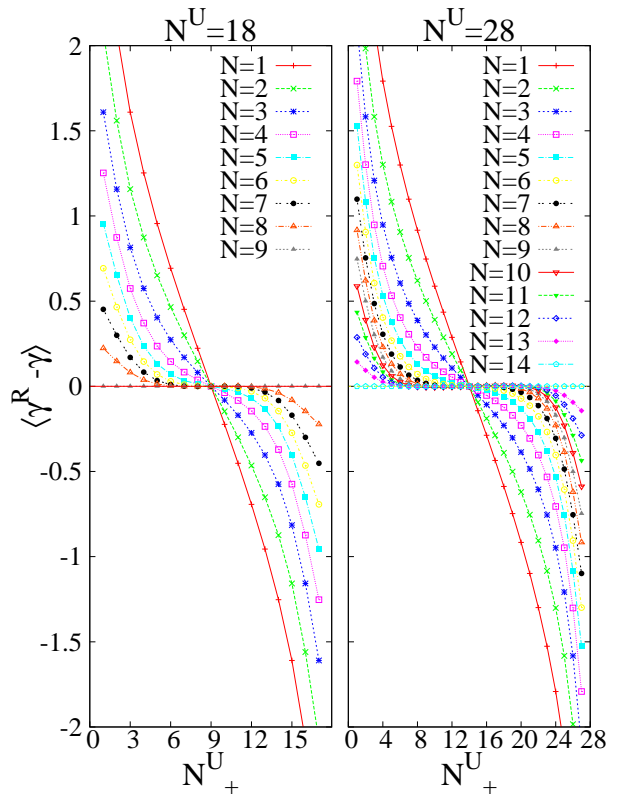,width=8.0cm,clip=}
\end{center}
\caption{\label{FIG3}  Test of  the validity  of the  equality  of the
$\langle\gamma \rangle$ and $\langle  \gamma ^R \rangle$ as a function
of  $N^U_+$ for  a paramagnetic  system, using  the  prescription from
Eq.~(\ref{eq:Afinitatlog})}
\end{figure}
\begin{figure}[htb]
\begin{center}
\epsfig{file=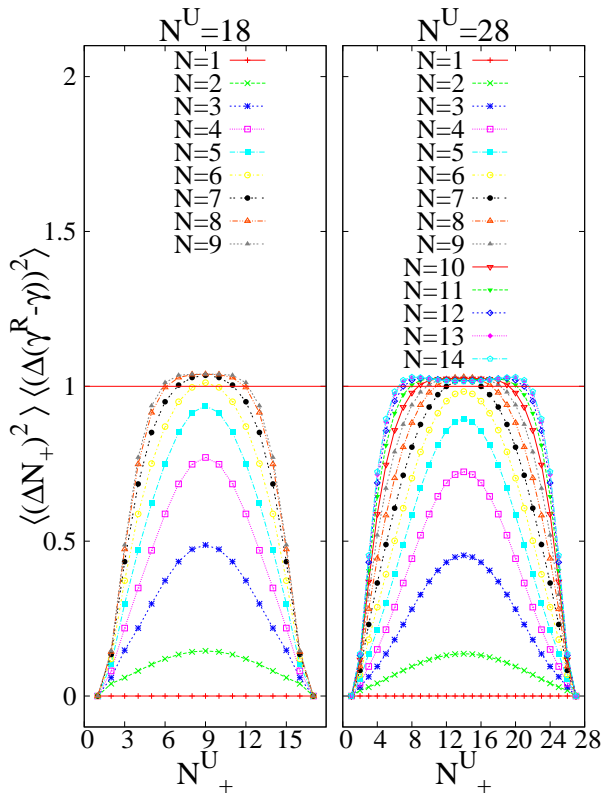,width=8.0cm,clip=}
\end{center}
\caption{\label{FIG4} Test of the validity of the uncertainty relation
as  a  function  of  $N^U_+$  for a  paramagnetic  system,  using  the
prescription from Eq.~(\ref{eq:Afinitatlog}).   The red line shows the
proposed lower bound.}
\end{figure}
For  illustrative  purposes, in  this  first  example  we compute  the
probability    for    the    system    to   reach    boundary    states
$p(N_+^{min})+p(N_+^{max})$.  Fig.~\ref{FIG2} shows  the value of this
probability  for two  universes of  sizes $N^U=18$  and $N^U=28$  as a
function of $N^U_+$  and for different values of  the system size $N$.
This result shows  that the probability of the  boundary states departs
from zero  when the  system or  reservoir are too  small and  when the
value of $N_+^U$ is too close to its extremes and this fact constrains
the number of available states in the system.

So  far there  has  been no  need  to take  one  of the  prescriptions
described  in  section \ref{subsec:Discrete}.   However,  in order  to
perform the  study of the  conjugated intensive variables we  must now
choose one of the prescriptions.  The first proposed option, according
to  Eq.~(\ref{eq:Afinitatlog}),   enables  us  to   benefit  from  the
separation of  the entropy density. As the  spins are non-interacting,
when we define the entropy  using the logarithm and neglect irrelevant
constants, it can be trivially separated as:
\begin{equation}
S(N_+) = \ln p(N_+)= S(N_+) + S^R (N^U_+-N_+)
\end{equation}
with
\begin{equation}
S(N_+) = \ln \frac{N!}{N_+! (N-N_+)!}
\end{equation}
and
\begin{equation}
\begin{split}
S^R(N^U_+-N_+)      \\     =      \ln     \frac{(N^U-N)!}{(N^U_+-N_+)!
 (N^U-N-N^U_++N_+)!}
\end{split}
\end{equation}
\begin{figure}[htb]
\begin{center}
\epsfig{file=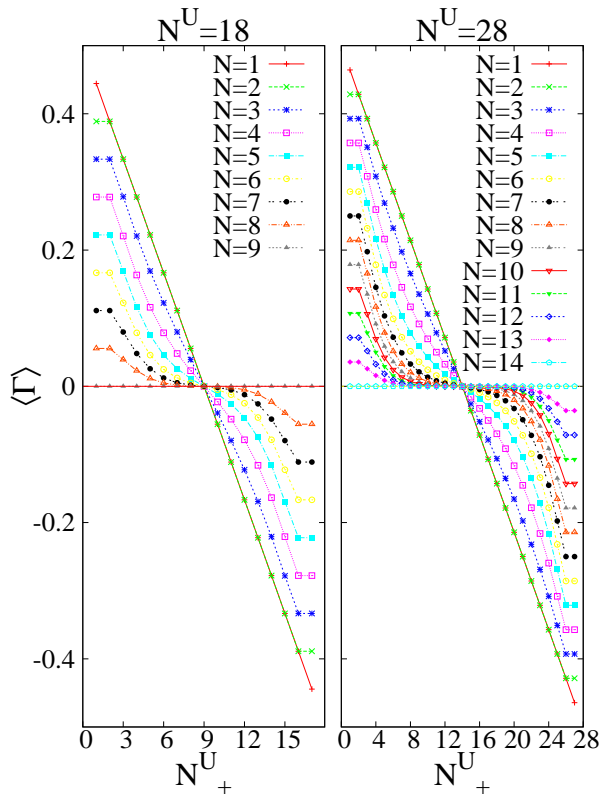,width=8.0cm,clip=}
\end{center}
\caption{\label{FIG5}  Test  of  vanishing  of  the  average  affinity
$\langle  \Gamma  \rangle$  for   a  paramagnetic  system,  using  the
prescription in Eq.~(\ref{eq:Afinitatdisc}) }
\end{figure}
\begin{figure}[htb]
\begin{center}
\epsfig{file=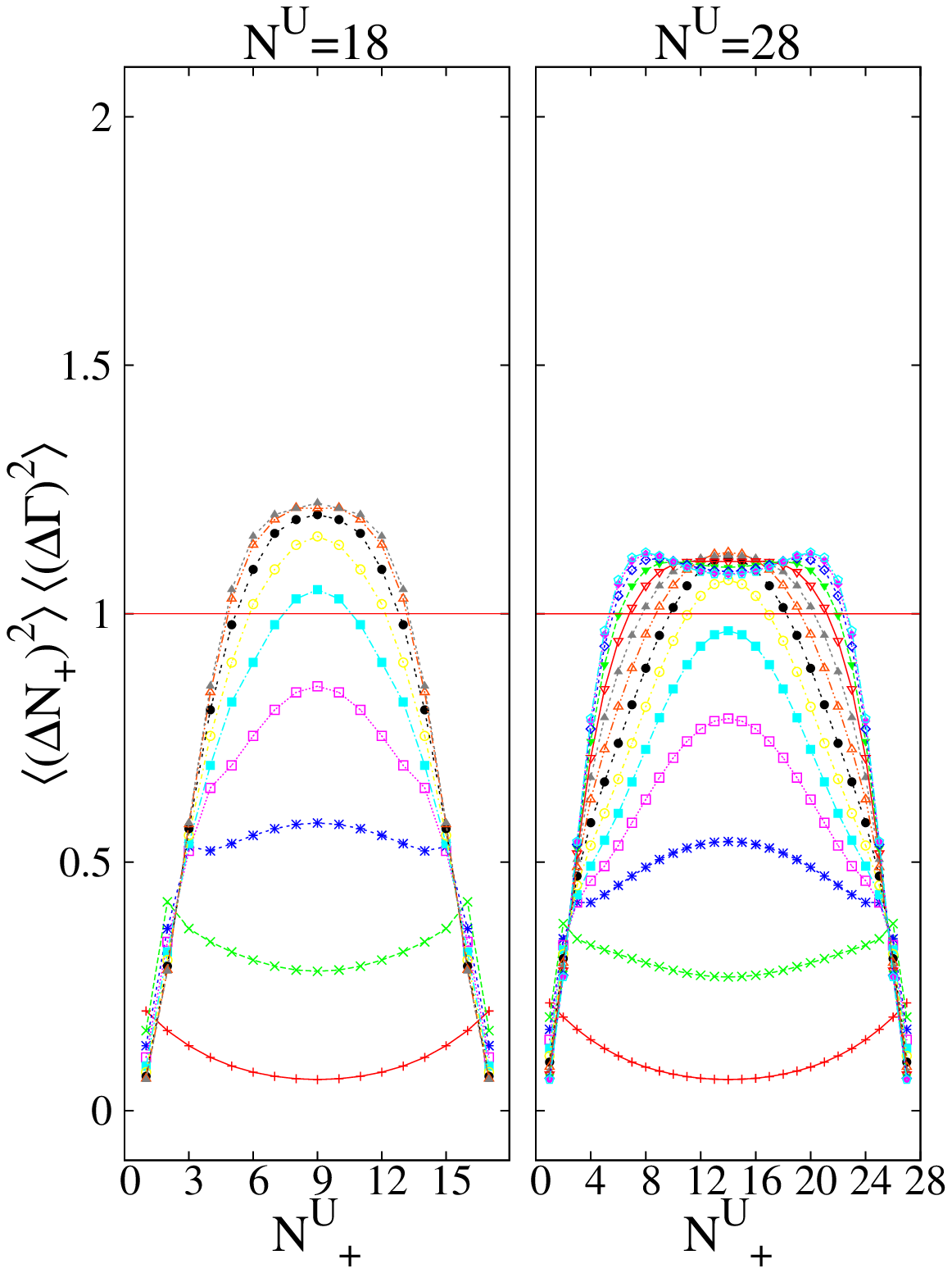,width=8.0cm,clip=}
\end{center}
\caption{\label{FIG6} Test of the validity of the uncertainty relation
for a paramagnetic  system using Eq.~(\ref{eq:Afinitatdisc}).  The red
line shows  the proposed lower  bound.  The symbols correspond  to the
values of $N$ indicated in the legends of the previous figures.}
\end{figure}
To  estimate the  intensive  variables of  the  system ($\gamma$)  and
reservoir ($\gamma^R$) we approximate the derivatives of $S$ and $S^R$
by a finite  difference scheme.  Since the variable  $N_+$ is bounded,
we  will use central  differences everywhere  except at  the extremes,
where we will use forward or backwards differences.  Thus:
\begin{equation}
\begin{split}
\gamma=\frac{d S}{d  N_+} \simeq \frac{S(N_++1)-  S(N_+-1)}{2} \\= \ln
\sqrt{\frac{(N-N_++1)(N-N_+)}{(N_++1)N_+}} \; ,
\end{split}
\end{equation}
and
\begin{equation}
\begin{split}
\gamma^R=\frac{d   S^R}{d  N^R_+}=-\frac{d   S^R}{d   N_+}  \simeq   -
\frac{S^R(N_++1)-        S^R(N_+-1)}{2}         \\        =        \ln
\sqrt{\frac{(N^U-N-N^U_++N_++1)(N^U-N-N^U_++N_+)}{(N^U_+-N_++1)
(N^U_+-N_+)}} \; .
\end{split}
\end{equation}
Fig.~\ref{FIG3}  shows  the  difference  $\langle  \gamma^R  -  \gamma
\rangle$ as a function of $N^U_+$, for universes of sizes $N=18$ (left
panel)  and  $N=28$  (right  panel)  and different  system  sizes  $N$
(ranging from $1$ to $N^U/2$). Notice that the roles of the system and
reservoir  are,  in  this  case,  symmetric.   As  can  be  seen,  the
difference  $\langle\gamma^R-\gamma\rangle$ vanishes  when $N_{+}^{U}$
(which determines the  available energy) is not close  to its upper or
lower extremes. It  is clearly observed that the  range of validity of
the result in Eq.~(\ref{eq:Equilibri2}) increases when the system size
$N$  approaches half  the universe  size $N^U/2$,  since we  have more
available states and therefore the importance of the boundary states on
the computed averages decreases.

Fig.~\ref{FIG4}   shows   the  test   of   the  uncertainty   relation
(\ref{eq:Resultatest2}) as a function of  the total number of up spins
$N^U_+$.   For   $N^U=18$  and  $N^U=28$,  and   different  values  of
$N$. Similar  conclusions to those observed in  Fig.~\ref{FIG3} can be
deduced: when $N$  and $N^U_+$ are far from  their boundary values, the
uncertainty relation is fulfilled.  Only when the fluctuating variable
$N_+$ is strongly constrained to a tiny number of states by the system
size  $N$ or  the  available  fixed universe  energy  $N^U_+$, do  the
surface terms neglected in  the derivation of the uncertainty relation
play a role and the bound is broken.

Alternatively, we  can study the  second method proposed  for discrete
systems   and    define   the   joint   affinity    $\Gamma$   as   in
Eq.~(\ref{eq:Afinitatdisc}).   In  this   case,  however,  it  is  not
possible to  separate $\Gamma$ into  two terms,
one corresponding to  the system and another one  corresponding to the
reservoir.  Numerical  computation for  the same cases  studied before
allows  us   to  plot  the  results  shown   in  Figs.~\ref{FIG5}  and
\ref{FIG6}.  The  same qualitative results concerning  the validity of
(\ref{eq:Equilibri}) and (\ref{eq:Resultatest}) can be concluded. Note
that  the values  of  $\langle\Gamma \rangle$  in Fig.~\ref{FIG5}  are
smaller  than  those  in  Fig.~\ref{FIG3}.   Therefore,  for  discrete
systems, the  choice of the approximation method  discussed in section
\ref{subsec:Discrete} may affect the results from a quantitative point
of view, even in such a simple case.

\subsection{\label{subsec:Example2} Ferromagnetic Ising system}

In this  second example we  study the case  of a discrete  system with
interactions. Since interaction can  induce correlations, in this case
we should not  assume that the entropy density  $S^U$ is separable and
thus we  should restrict  ourselves to only  testing the  validity for
equations (\ref{eq:Equilibri}) and (\ref{eq:Resultatest}).
\begin{figure}[htb]
\begin{center}
\epsfig{file=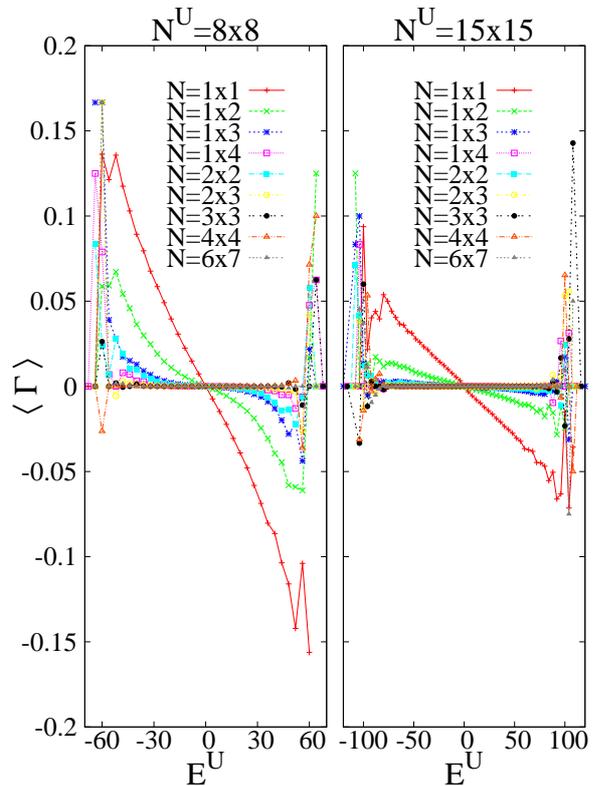,width=8.0cm, clip=}
\end{center}
\caption{\label{FIG7}     $\left\langle\Gamma\right\rangle$    plotted
against the energy of the universe for a 15x15 and a 8x8 universe, and
for various  sizes of the corresponding  systems.  As the  size of the
system  increases, the  expected value  of $\Gamma$  goes to  zero, as
expected.}
\end{figure}
\begin{figure}[htb]
\begin{center}
\epsfig{file=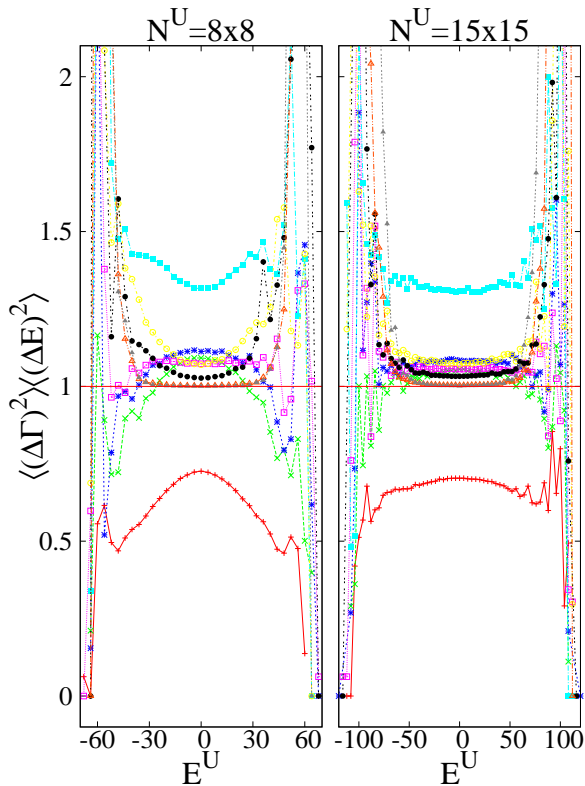,width=8.0cm, clip=}
\end{center}
\caption{\label{FIG8}  Test for  the fluctuation  theorem  in discrete
systems.  Symbols  are the  same as in  the previous figure. As  in figure
\ref{FIG2}, the red line shows the proposed lower bound.}
\end{figure}
Let us  consider a  universe consisting  of an Ising  model on  a two-
dimensional  square lattice  with  size $N^U  =  L^U\times L^U$,  with
nearest-neighbour  interactions and periodic  boundary conditions.  The
system of interest will be a  rectangular subset of spins with size $N
=  L_x \times  L_y$.  Even  though  a general  treatment requires  the
simultaneous analysis of energy and magnetization fluctuations, in the
absence of  an external field, it is  enough to analyse  only the energy.
The Hamiltonian of the universe reads
\begin{equation}
{\cal H}^U = -\sum_{i,j}^{n.n.} s_i s_j \; ,
\end{equation}
where $s_i=\pm 1$ are the spin  variables and the sum extends over all
the  $2N^U$ nearest-neighbour pairs  of the  universe. Note  that the
energy of the universe will be discretized in steps $\delta E^U=4$ and
ranges  from  $-2N^U$  to  $2N^U$  For any  given  configuration,  the
evaluation of  the energy  of the system  will be done  by considering
that the  energy associated with the  bonds along the  boundary of the
system is  equally divided  between the system and  reservoir.  Therefore,
the energy of the system is discretized in steps $\delta E=1$.

In  order to  perform a  microcanonical  analysis of  the averages  at
constant $E^U$ we have used  the following procedure. We have randomly
generated  spin configurations  and classified  them according  to the
universe energy.   Inside the set  corresponding to the same  value of
$E^U$ we have computed  the fraction of configurations that correspond
to every possible value of the  energy $E$ of the system and performed
a statistical estimation of its probability $p(E)$.

It is important to note that, because of the finite size of the random
sample   of  states   that  we   have  generated   ($3   \times  10^7$
configurations), we do not obtain  states with an {\it a priori} small
but non-zero probability. This means that there is a certain number of
possible values of $E$ for which we wrongly estimate zero probability.
As  the logarithmic  prescription (\ref{eq:Afinitatlog})  discussed in
\ref{subsec:Discrete}  would  give  problems  with infinities  at  the
neighbouring values, we have only used the second prescription given by
Eq.~(\ref{eq:Afinitatdisc}).  Even  in this  second case, in  order to
avoid problems with  the computation of the variance  of $\Gamma$ (for
which  a  $p_i$ survives  in  the  denominator),  the states  with  an
estimated zero probability have been  excluded: it is as if they could not be
reached by the system.

Figures  \ref{FIG7}  and  \ref{FIG8}  show the  results  obtained  for
simulations  of a  $8 \times  8$ and  a $15  \times 15$  universe, for
various sizes of the  system.  These are qualitatively compatible with
those    obtained   for   the    system   without    interactions   in
Sec.~\ref{subsec:Example1}.   The uncertainty relations  are fulfilled
except  in the  very extreme  case in  which the  number  of available
states of  the system and/or reservoir  is so small  that the boundary
effects are dominant.

\section{\label{sec:Conclusions}Discussion and Conclusions}

We  have  shown  that  uncertainty  relations can  be  established  in
statistical mechanics  when considering a universe  (isolated from the
outer world)  separated into two  parts, called the system  and reservoir,
which  are  in thermodynamic  equilibrium  and  interact by  exchanging
"mechanical  properties" such  as energy,  volume, particles,  ...  If
$X_i$ is one of these  properties corresponding to the system, we have
found that the following  inequality must be satisfied, $\sqrt{\langle
(\Delta X_i)^2  \rangle \langle (\Gamma_i)^2 \rangle}  \geq 1$.  Here,
$\Gamma_i$  is  an affinity  conjugated  to  the  variable $X_i$  which
determines the departure from strict equilibrium of actual microscopic
configurations  of  the two  interacting  systems  and which  satisfies
$\langle  \Gamma_i  \rangle  =  0$.   Furthermore,  when  the  entropy
satisfies  separability,   these  affinities  simply   reduce  to  the
difference   $\gamma_i^R  -  \gamma_i$   of  the   intensive  variable
conjugated to $X_i$, in the reservoir and the system.
\begin{figure}[ht]
\begin{center}
\epsfig{file=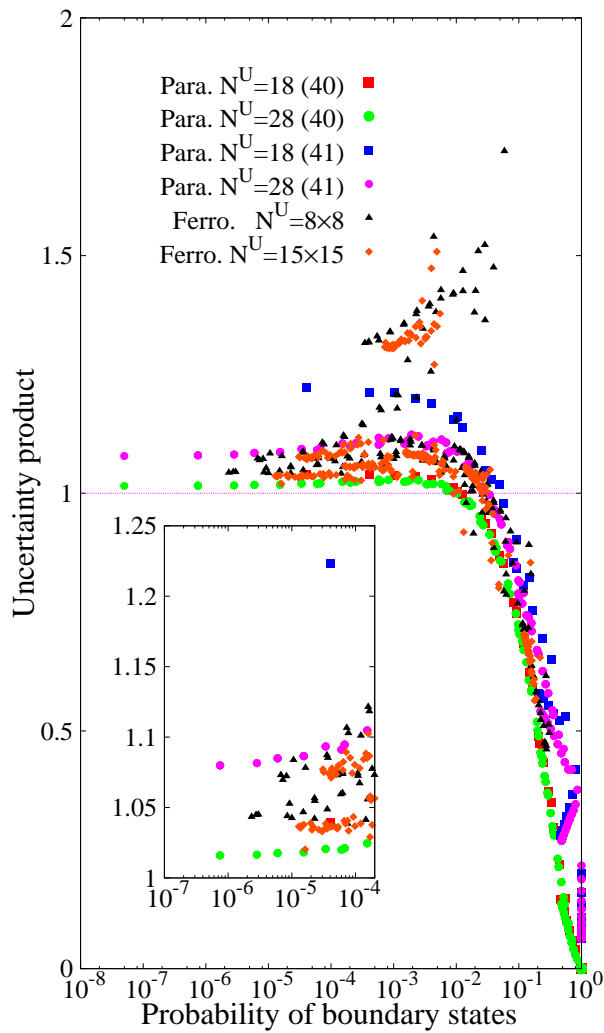,width=8.0cm, clip=}
\end{center}
\caption{\label{FIG9} Uncertainty  product against the  probability of
boundary states for the paramagnetic Ising system and the ferromagnetic
Ising model.   Data includes  different universe sizes,  system sizes,
and different  values of the  mechanical variables of the  universe. We
have also included the data corresponding to the two different methods
proposed for  the evaluation of  the affinities for  discrete systems,
given  by   Eqs.~(\ref{eq:Afinitatlog})  and  (\ref{eq:Afinitatdisc}),
indicated in the legend by  the numbers between parenthesis. The inset
shows a magnification of the  region above $1$ for probabilities below
$2\times 10^{-4}$.}
\end{figure}
Assuming  that  the  intensive  variable  of the  reservoir  does  not
fluctuate  as  expected  for   very  large  reservoirs,  the  proposed
inequality reduces to the more standard uncertainty relation involving
only system variables
\begin{equation}
\sqrt{\langle (\Delta X_i)^2 \rangle \langle (\gamma_i)^2 \rangle} \ge
1
\end{equation}
Furthermore, assuming  that the fluctuations in the  system are small,
by expanding  to first  order, the uncertainty  relation leads  to the
following equality:
\begin{equation}
\langle (\Delta X_i)^2  \rangle = \frac{1}{\langle (\Delta (\gamma_i^R
  - \gamma_i))^2 \rangle}
\end{equation}
It   is  interesting   to  point out,  than   within  this   first-order
approximation, when  the relevant conjugated variables  are the energy
and  the inverse  temperature, combining  Eqs.~(\ref{eq:Lindhart}) and
(\ref{eq:Lindhart4}), one gets:
\begin{equation}
\sqrt{\langle (\Delta E)^2 \rangle \langle (\Delta \beta)^2 \rangle} =
C^R/(C + C^R) \leq 1 \; ,
\end{equation}
where  $C$  and  $C^R$ are  the  heat  capacities  of the  system  and
reservoir.   Therefore,  in this  last  case  fluctuations of  inverse
temperature are not, in general, expected to be proportional to energy
fluctuations  \cite{Chui1992}. Only  in the  limit $C/C^R  \ll  1$ is the
standard  result  $\sqrt{\langle(\Delta  E)^2  \rangle  \langle(\Delta
\beta)^2\rangle} = 1$ recovered.

The derivation of  the above results has required  the assumption that
the probability for the boundary states  (on the boundary of the set of
mechanical  variables $\Omega$) vanishes.   This assumption,  which is
reasonable for  universes described  by continuum variables,  might be
incorrect for the discrete case,  when universes are small or too 
constrained.

We have numerically  studied the validity of the  above results in two
different  mechano-statistical lattice  systems: the  Ising paramagnet
and  the  Ising  ferromagnet.   We  have found  that  the  uncertainty
relations hold except  in the case in which the  boundary states have a
high  probability  of occurring.   In  particular,  from  the examples  in
section \ref{sec:Examples},  we can  numerically evaluate a  bound for
this  probability.   Fig.  \ref{FIG9}  shows  the uncertainty  product
$\langle  (\Delta X_i)^2 \rangle  \langle \Gamma_i^2  \rangle$ against
the probability of  the boundary states for the  Ising paramagnet (with
the two  possible definitions of  the affinities for  discrete systems
given by Eqs. ~(\ref{eq:Afinitatlog}) and (\ref{eq:Afinitatdisc})) and
the Ising ferromagnet.  The figure clearly shows that the breakdown of
the uncertainity relation  may occur only when the  probability of the
boundary states is above 0.01.  In all the other cases, the uncertainty
product is  larger than one and, in  fact, tends to 1  from above when
the size  of the universe  increases, as can  be seen in the  inset of
Fig.\ref{FIG9}.

These  results  are  relevant  from  an experimental  point  of  view.
Focusing on the case of the  energy $E$ and the inverse of temperature
$\beta$ as conjugated variables, we  can identify the reservoir with a
given  body and the  system with  the thermometer  used to  measure its
temperature. Commonly, the thermometer is chosen so that it has a heat
capacity much  smaller than the  heat capacity of the  body.  However,
for small  bodies, when the heat  capacity of the  thermometer and the
body  are comparable,  the results  derived  here must  be taken  into
account: (i) the inverse  temperature $\langle \beta \rangle$ measured
by the thermometer con strongly fluctuate, but still its average value
must be identified with the  average of the inverse temperature of the
body  $\langle \beta^R  \rangle$;  and (ii)  the  fluctuations of  the
difference   $\beta^R-\beta$   should  be   related   to  the   energy
fluctuations through the proposed uncertainty relation.

Future extensions of this work  will deal with the analysis of systems
described by continuous variables,  where direct counting of states is
not straightforward. The results will be presented elsewhere.

\vspace{5mm}

\begin{acknowledgments}
This work has received financial  support from the Spanish Ministry of
Innovation  and Science (project  MAT2010-15114). G.T.  acknowledges a
grant  from the  University  of  Barcelona. We  thank  A.Saxena for  a
critical reading of the manuscript.
\end{acknowledgments}

\end{document}